\documentstyle[12pt,aps,floats,epsf]{revtex}

\begin{document}

\begin{flushright}
JLAB-THY-00-26 \\
\end{flushright}

\vspace{0.5cm}

\centerline{\bf Renormalons as dilatation modes in the functional space }

\vspace{1cm}
\centerline{A. BABANSKY and I. BALITSKY}
\vspace{0.5cm}
\centerline{ Physics Department, Old Dominion University, Norfolk 
VA 23529} 
\centerline{and}
\centerline{Theory Group, Jefferson Lab, Newport News VA 23606}
\centerline{e-mail: \underline{balitsky@jlab.org}  and \underline{babansky@jlab.org}}  
\vspace{1cm}

\centerline{\bf Abstract} 

\vspace{0.4cm}
There are two sources of the factorial large-order behavior of a typical
perturbative series.  
First, the number of the different Feynman diagrams may
be large; second, there are abnormally large diagrams known as
renormalons.  It is well known that the large combinatorial number of diagrams 
is described by instanton-type solutions of the
classical equations.  We demonstrate that from the
functional-integral viewpoint the renormalons do not correspond to a particular
configuration but manifest themselves as dilatation modes in the functional
space.

\newpage
\narrowtext
\section{Introduction}
\bigskip
It is well known  that the 
perturbative series in a typical quantum field theory is at best
asymptotic: the coefficients in front of a typical perturbative 
expansion grow like $n!$ where n is the order of the perturbation series 
(see e.g. the book \cite{zinn}). 
There are two sources of the $n!$ 
behavior which correspond to two different situations. In the first case
all Feynman diagrams are $\sim 1$ but their number is large ($\sim n!$)
\cite{lip1}. 
In the second case, we have just 
one diagram but it is abnormally big -- $\sim n!$ (the famous `t Hooft 
renormalon \cite{thoft1}). The first type of factorial behavior is not specific 
to a field 
theory; for example, 
it can be studied in the anharmonic-oscillator quantum mechanics. On the 
contrary, renormalon singularities can occur only in field theories with
running coupling constant (for the review of the renormalons, 
see ref.\cite{beneke} and references therein). 

It is convenient to visualize the large-order behavior of perturbative series
 using the 't Hooft picture of singularities\cite{thoft1}. 
 Consider Adler's function related to the
 polarization operator in QCD
 \begin{eqnarray}
 &&\hspace{-3cm}D(q^2)={4\pi^2}q^2{d\over dq^2}{1\over 3q^2}\Pi(q^2),\label{fla1}\\
&&\hspace{-3cm}\Pi(q^2)=\int\! dx e^{iqx}\int\! D\bar{\psi}D\psi DA 
j_{\mu}(x)j_{\mu}(0)e^{-S_{\rm QCD}}
 \nonumber
\end{eqnarray}
Suppose we write down $D(q^2)$ as a Borel integral
\begin{equation}
\hspace{-3cm}D(\alpha_s(q))=\int_0^{\infty}dt D(t)e^{-{4\pi \over \alpha_s(q)}t} 
\label{fla2}
\end{equation}
The divergent behavior of the original series $D(\alpha_s(q)$ is encoded
in the singularities of its Borel transform shown in Fig.1. 
\begin{figure}[htb]
\mbox{
\epsfxsize=12cm
\epsfysize=2cm
\hspace{1cm}
\epsffile{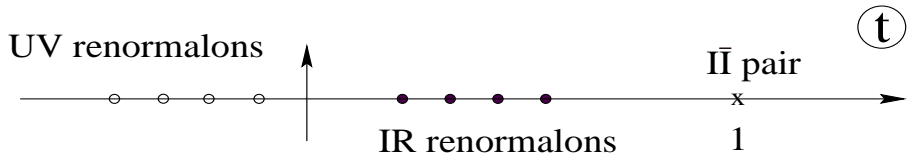}}
\vspace{0cm}
{\caption{ 't Hooft picture of singularities in the Borel plane for QCD. 
\label{fig1}}}
\end{figure}
In QCD, we have 
two types of 
singularities: renormalons (ultraviolet or
infrared) and instanton-induced singularities \footnote{
Actually, the instanton itself is not related to the 
divergence of 
perturbative series
since it belongs to a different topological sector. The first topologically 
trivial classical
configuration, which contributes to the divergence of perturbation 
theory, is a weakly coupled instanton-antiinstanton
pair \cite{bogfat}.
}.
The ultraviolet (UV) renormalons are located at
$t=-{1\over b},{2\over b},{3\over b}...$ ($b=11-{2\over 3}n_f$). 
In terms of Feynman diagrams they come from the regions of hard momenta 
in renormalon bubble chain, see Fig. 2.
\begin{figure}[htb]
\hspace{5cm}\vspace{0cm}
\mbox{
\epsfxsize=12cm
\epsfysize=2.7cm
\hspace{2cm}\vspace{-2cm}
\epsffile{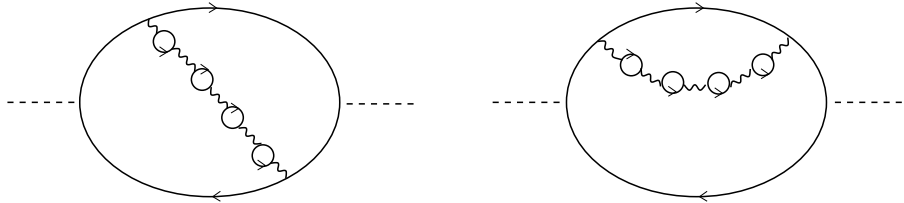}}
\vspace{0cm}
{\caption{ Typical renormalon bubble chain diagrams for the polarization
operator in QCD. 
\label{fig1a}}}
\end{figure}
The infrared (IR) renormalons placed at
$t={2\over b},{3\over b}...$ come from the region of 
soft momenta in the bubble chain diagrams. The instanton 
singularities are located at 
$t=1,2,3,...$ and they correspond to the large number of graphs 
which are 
summed up to the contribution of $I\bar{I}$ configurations 
in the functional space. 
The main result of this paper is that unlike the instanton-type singularities, 
the renormalons do not correspond
to a particular configuration  but manifest 
themselves as dilatation modes in the functional space. 
 
 The paper is organized as follows. In Section 2 we relate Borel representation
to the functional integral along the instanton-antiinstanton valley in 
the double-well quantum mechanics. In Section 3 we describe the conformal 
valley in QCD and discuss the relation between valleys and Borel 
representation in the field theories with running coupling constant. 
Sect. 4 and 5 are devoted to the interpretation of IR and UV
renormalons as dilatation modes of the valley. In the last section we discuss 
general aspects of our approach and outline the direction of the future work.

\section{Valleys and Borel plane in quantum mechanics}
The interpretation of renormalons as dilatation modes is based upon the
similarity of the functional integral in the vicinity of a valley in the
functional space\cite{byng} to the Borel representation (2).  
At first, we will consider the
quantum  mechanical example without renormalons and then demonstrate that in a
field theory the same integral along the valley leads to renormalon
singularities. With QCD in view, we consider the double-well 
anharmonic oscillator described by the functional
integral  
\begin{equation} Z~=\int\! D\phi e^{-{S(\phi)\over g^2}}
\label{fla3} 
\end{equation}
where
\begin{equation}
S(\phi)=\int\! dt\left({\dot{\phi}^2\over 2}+{\phi^2(1-\phi)^2)\over 2}
\right)
\label{fla3a} 
\end{equation}
The large-order behavior in this model is determined by 
the instanton-antiinstanton ($I\bar{I}$)
configuration. The $I\bar{I}$ valley for the
double-well system may be chosen as 
\begin{equation}
f_{\alpha}(t-\tau)={1\over 2}\tanh{t-\tau+\alpha\over 2}-
{1\over 2}\tanh{t-\tau-\alpha\over 2}
\label{fla4}
\end{equation}
It satisfies the valley equation\cite{yung}\cite{bshafer}  
\begin{equation}
{12\over \xi^2}w_{\alpha}(t)f'_{\alpha}(t)=
L_{\alpha}(t) 
\label{flav}
\end{equation}
where $\xi\equiv e^{\alpha}$,
$f'_{\alpha}\equiv{\partial\over\partial\alpha}f_{\alpha}$, and
$L_\alpha(t)=\left.{\delta S\over
\delta\phi}\right|_{\phi=f_{\alpha}(t)}$. Here 
 $
 w_\alpha(t)={\xi\over 4}\sinh\alpha(\cosh\alpha\cosh t+1)^{-1}
 $ is
the measure in the functional space so $(f,g)\equiv\int
dtw_\alpha(t)f(t)g(t)$. The valley (\ref{fla4}) connects two classical
solutions: the perturbative vacuum at $\alpha=0$ and the infinitely separated
$I\bar{I}$ pair at $\alpha\rightarrow \infty$. The $I\bar{I}$ separation
$\alpha$ and the position of the $I\bar{I}$ pair $\tau$ are the two collective
coordinates of the valley. In order to integrate over the small
fluctuations near the configuration (\ref{fla4}), we  
introduce 
two $\delta$-functions
$\delta(\phi(t)-f_{\alpha}(t-\tau),\dot{f}_{\alpha}(t-\tau))$ and 
$\delta(\phi(t)-f_{\alpha}(t-\tau),{f}'_{\alpha}(t-\tau))$
restricting the integration
along the two collective coordinates. Following the standard procedure 
(Faddeev-Popov trick), we 
insert 
\begin{equation}
1=\int d\alpha\left[(f'_{\alpha},f'_{\alpha})+(\phi-f_{\alpha},f'_{\alpha}+
{w'\over w}f_{\alpha})\right]\delta(\phi-f_{\alpha},f'_{\alpha})
\label{edin}
\end{equation}
and
\begin{equation}
1=\int d\tau\left[(\dot{f}_{\alpha},\dot{f}_{\alpha})+(\phi-f_{\alpha},
\dot{f}_{\alpha}+
{\dot{w}\over w}f_{\alpha})\right]\delta(\phi-f_{\alpha},\dot{f}_{\alpha})
\end{equation}
in the functional integral (\ref{fla3}).
Next steps are the shift
$\phi(t)\rightarrow\phi(t)+f_{\alpha}(t-\tau)$,  expansion in quantum deviations
$\phi(t)$ and the gaussian integration in the first nontrivial order
in perturbation theory. After the shift, the the linear
term in the exponent $\int dt \phi(t)L_\alpha(t)$ is disabled 
due to the valley equation (\ref{flav}) 
(recall $\delta(\phi,f'_{\alpha})$ in the integrand coming from 
eq. (\ref{edin})) so the functional
integral (\ref{fla3}) for the vacuum energy 
in the leading order in $g^2$ 
reduces to
\begin{eqnarray} &&T\int\!
d\alpha\!\int\! D\phi 
(f'_{\alpha},f'_{\alpha})(\dot{f}_{\alpha},\dot{f}_{\alpha})
\delta(\phi,f'_{\alpha})\delta(\phi,\dot{f}_{\alpha})
e^{-{1\over g^2}
\left[S_\alpha+{1\over 2}\int\! dt\phi(t)\Box_{\alpha}\phi(t)\right]} 
+O(g^2).
\label{fla5}
\end{eqnarray}
Here T  (the total "volume" in one space-time dimension) is the result
of trivial integration over $\tau$,  
$\Box_{\alpha}=-\partial^2 +1-6f_{\alpha}(1-f_{\alpha})$ is the operator of
second derivative of the action and
\begin{equation}
S_{\alpha}\equiv S(\xi)={6\xi^4-14\xi^2\over(\xi^2-1)^2}-{17\over 3}+
{12\xi^2+4\over(\xi^2-1)^3}\ln\xi
\label{fla6}
\end{equation}
is the action of $I\bar{I}$ valley. 
Performing Gaussian integrations we get
\begin{eqnarray}
&&\hspace{-2cm}T{1\over g^2}\int_0^{\infty}\! d\alpha e^{-{1\over g^2}S_\alpha}F(\alpha),
\label{fla7}\\
&&\hspace{-2cm}F(\alpha)={(\det\Box_{\alpha})^{-1/2}(f'_{\alpha},f'_{\alpha})
(\dot{f}_{\alpha},\dot{f}_{\alpha})\over
\left(f'_{\alpha},\Box^{-1}_{\alpha}w_{\alpha}f'_{\alpha}\right)^{1/2}
\left(\dot{f}_{\alpha},\Box^{-1}_{\alpha}
w_{\alpha}\dot{f}_{\alpha}\right)^{1/2}}.
\label{fla8}
\end{eqnarray}
At $\alpha\rightarrow \infty$ we have the widely separated $I$ and $\bar{I}$. 
In this case, the determinant of the $I\bar{I}$ configuration
factorizes into a product of two
one-instanton determinants (with zero modes excluded) so that
$F(\alpha)\rightarrow {\rm const}$ at $\alpha\rightarrow \infty$. The
divergent part of the integral (\ref{fla7}) corresponds to the second
iteration of the one-instanton contribution to the vacuum energy, therefore
it must be subtracted from the $I\bar{I}$ contribution to $E_{\rm vac}$:
\begin{equation}
g^2E_{\rm vac}(g^2)=\int\! d\alpha\!
\left(e^{-{1\over g^2}S_\alpha}F(\alpha)-e^{-{1\over g^2}2S_I}F(\infty)\right)
\label{fla9}
\end{equation}
where $S_I$ is the one-instanton action. Since $S_\alpha$ is a monotonous
function of $\alpha$ one can invert Eq. (\ref{fla6}) and obtain
\begin{equation}
g^2E_{\rm vac}(g^2)=\int_0^{2S_I}\! dS
e^{-{1\over g^2}S}F(S)\left({1\over S-2S_I} \right)_+
\label{fla10}
\end{equation}
which has the desired form of the Borel integral (\ref{fla1}) for the vacuum
energy. Thus the leading singularity for $E_{\rm vac}(S)$ is
$F(2S_I)/(S-2S_I)$. Note that our semiclassical calculation
does not give the whole answer for the Borel transform of vacuum
energy ($F(S)\not\! =E_{\rm vac}(S)$ in general) since we threw away the higher
quantum fluctuations around the $I\bar{I}$ pair in Eq. (\ref{fla5}) yet it
determines the the leading singularity in the Borel plane.

\section{$I\bar{I}$ valley and Borel integral in QCD}
The situation with the instanton-induced asymptotics of perturbative series in
a field theory such as QCD is pretty much similar to the case of quantum
mechanics with one notable exception:
in QCD there is an additional dimensional parameter -- 
the overall size of 
the $I\bar{I}$ configuration $\rho$. The classical $I\bar{I}$ action does not depend
on this parameter but the quantum determinant does, leading to the
replacement 
\begin{equation}
e^{-{1\over g^2}S}\rightarrow e^{-{1\over g^2(\rho)}S}
\label{fla11}
\end{equation}
so the  Borel integrand 
have the following generic form
\begin{equation}
F(S)\sim\int d\rho e^{-{1\over g^2(\rho)}S} {\cal F}(\rho)
\label{fla12}
\end{equation}
The divergence of this integral at either large or small $\rho$ determines
the position of singularities of $F(s)$. We will demonstrate (by purely
dimensional analysis) that
 ${\cal F}(\rho)\sim \rho^{-5}$ at $\rho\rightarrow \infty$ 
and
${\cal F}(\rho)\sim \rho$ at $\rho\rightarrow 0$  
leading to the IR renormalon at $S={32\pi^2\over b}$ and the UV
renormalon at $S=-{16\pi^2\over b}$, respectively
\footnote{In its present form, our analysis of renormalons is applicable 
to the ``off-shell'' processes which can be related to the Euclidean 
correlation functions of two (or more) currents.  An example of a
different type is the IR renormalon in the pole mass of a heavy quark
located at $S={16\pi^2\over b}$\cite{ural}}.

The $I\bar{I}$ valley in QCD\cite{yung} can be chosen as a 
conformal transformation of
the spherical configuration
\begin{equation}
A_{\mu}(x)=-i{\bar{\sigma}_{\mu}x-x_{\mu}\over x^2}f_{\alpha}(t)
\label{fla13}
\end{equation}
with $t=\ln x^2/d^2$ where $d$ is an arbitrary scale. (We use the notations
$x\equiv x_{\mu}\sigma_{\mu},~~\bar{x}\equiv x_{\mu}\bar{\sigma}_{\mu}$ 
where
$\sigma_{\mu}=(1,-i\vec{\sigma}),~~\bar{\sigma}_{\mu}=(1,i\vec{\sigma})$). 
To obtain the $I\bar{I}$ configuration with arbitrary sizes $\rho_1,
\rho_2$ and separation $R$ one performs shift  $x\rightarrow x-a$, inversion
$x\rightarrow {d^2\over x^2}x$ and second shift $x\rightarrow x-x_0$ 
(see Fig. 3). 
\begin{figure}[htb]
\mbox{
\epsfxsize=10cm
\epsfysize=5cm
\hspace{2cm}
\epsffile{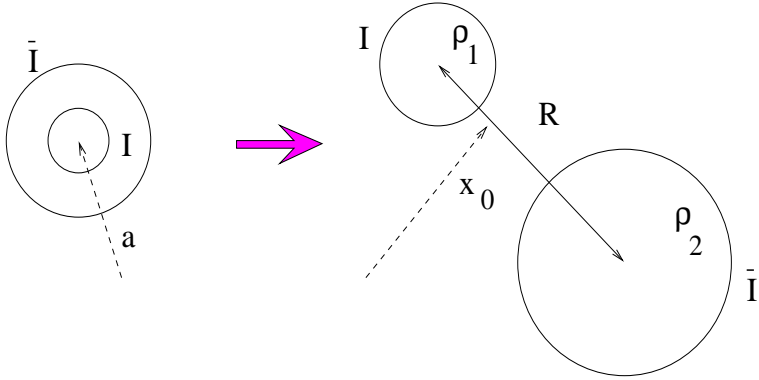}}
\vspace{0cm}
{\caption{ Conformal $I\bar{I}$ valley in QCD. 
\label{fig2}}}
\end{figure}
The
resulting valley is the sum of the $I$ and $\bar{I}$ in the singular
gauge in the maximum attractive orientation plus the additional term (which  
is small at large $I\bar{I}$ separations)
\begin{equation}
A^v_{\mu}(x)=A^I_{\mu}(x-x_0)+A^{\bar I}_{\mu}(x-x_0)+B_{\mu}(x-x_0)
\label{fla14}
\end{equation}
where 
\begin{eqnarray}
A^I_{\mu}(x)&=&-i\rho_1^2{\sigma_{\mu}\bar{x}-x_{\mu}\over
x^2(x^2+\rho_1^2)}\label{fla15}\\
A^{\bar I}_{\mu}(x)&=&
-i\rho_2^2{R\bar{\sigma}_{\mu}(x-R)\bar{R}-(x-R)_{\mu}
\over R^2(x-R)^2((x-R)^2+\rho_2^2)}. 
\nonumber
\end{eqnarray}
The explicit expression for $B_{\mu}$ can be found in \cite{bshafer}. 
The action of the $I\bar{I}$ valley 
(\ref{fla14}) is  equal to the action of the spherical configuration
(\ref{fla13}) given by (\ref{fla6}):
\begin{equation}
S^v(z)=48\pi^2S(\xi),~~ \xi=z+\sqrt{z^2-1},
\label{fla16}
\end{equation}
where the "conformal parameter" $z$ is given by
\begin{equation}
z=(\rho_1^2+\rho_2^2+R^2)/(2\rho_1\rho_2).
\label{fla17}
\end{equation}

Let us now calculate the polarization operator (\ref{fla1}) in the valley
background. The collective coordinates are the sizes of instantons $\rho_i$,
separation $R$, overall position $x_0$ and the orientation in the color
space (the valley of a general color orientation has the form 
$ {\cal O}_{ab}A^{v}_b$ where  ${\cal O}$ is an arbitrary $SU(3)$ matrix). 
The structure of Gaussian result for the polarization operator 
in the valley background 
is 
\begin{equation}
\int_0^{\infty}\!{d\rho_1d\rho_2\over\rho_1^5\rho_2^5} d^4R d^4x_0 
d{\cal O}  \Pi^v(q){1\over g^{17}}e^{-{S^v(z)\over g^2}}\Delta(\rho_i,R)
\label{fla18}
\end{equation}
where 17 is a number of the collective coordinates
\footnote
{At $z\rightarrow \infty$ (for weakly interacting $I$ and $\bar{I}$)
the mode corresponding to relative orientation in the color space
becomes non-gaussian so one should introduce an additional collective
coordinate corresponding to this quasizero mode. We are interested,
however, in $z\sim 1$ where this mode is still Gaussian so it is taken 
into account in quantum determinant $\Delta$.
}.
Here $\Pi^v(q)$ is a Fourier transform of
$$
\Pi^v(x)=
(\sum e_q^2){\rm Tr}\gamma_{\mu}G(x,0)\gamma_{\mu}G(0,x)
$$
where $G(x,y)$ is the Green
function in the valley background. 
The factor $\Delta(\rho_i,R)$ in Eq. (\ref{fla18}) is the quantum
determinant - the result of Gaussian integrations near the valley (\ref{fla14})
with the additional factors due to the restricted integrations along the
collective coordinates (cf. eq. (\ref{fla8})).
For our purposes, it is convenient to introduce the conformal parameter $z$ and
the average size $\rho=\sqrt{\rho_1\rho_2}$ as the collective coordinates in
place of  $\rho_1$ and $\rho_2$. We have 
\begin{eqnarray}
\int\! dz{d\rho\over\rho^9} d^4R d^4x_0 
\Pi^v(q){1\over g^{17}(\rho)}
F(z,R^2/\rho^2)e^{-{S^v(z)\over g^2(\rho)}}
\label{fla20}
\end{eqnarray}
where $F(z,R^2/\rho^2)$ contains 
$\theta(z-1-{R^2\over2\rho^2})$ (see Eq. (\ref{fla17})). 
We have included in $F$ the trivial integral over color orientation 
which gives the volume of $SU(3)$ group.

The main effect of the quantum determinant $\Delta$ is the replacement
of the bare coupling constant $g^2$ in Eq. (\ref{fla18}) by the effective
coupling constant $g^2(\rho)$ so the remaining function $F$ is the
(dimensionless) function of the ratio $R^2/\rho^2$ and the conformal parameter.
This is almost evident from the renormalizability of the theory since the only
dimensional parameters are $\rho$ and $R$. Formally, one can prove  
that rescaling of the configuration by factor
$\lambda$ (so that $\rho\rightarrow \lambda\rho$ and $R\rightarrow \lambda R$)
leads at the one-loop level to multiplication of the determinant 
by a factor $\lambda^{bS^v(z)/8\pi^2}$
due to conformal anomaly (see Appendix). 

\section{IR renormalon from the dilatation mode of the $I\bar{I}$ valley}

Consider the singularities of the integral (\ref{fla20}). The function
$F(z,R^2/\rho^2)$  is non-singular since the singularity
in $\Phi$ ($\equiv$ singularity in $\Delta$) would mean a non-existing zero
mode in quantum determinant. Moreover, the integration over $R$ is finite 
due to 
$\theta(z-1-{R^2\over2\rho^2})$ indicating that the only source of singularity 
at finite 
$z$ is the divergence of the $\rho$ integral at either large or small $\rho$.
(At $z\rightarrow\infty$, in a way similar to the derivation of Eq.
(\ref{fla10}) we obtain the first instanton-type singularity
located at $t=1$\cite{bal91}).    

Let us
demonstrate that the singularity of the integral
(\ref{fla20}) at large $\rho\gg{1\over q}$ corresponds to the IR renormalon.  
 The polarization operator 
$\Pi(q)$ in the background of the 
large-scale vacuum fluctuation reduces to\cite{svz} (see Fig. 3)
\begin{eqnarray}
&&\Pi(q)=\int d^4x e^{iqx}\langle j_{\mu}(x)j_{\mu}(0)\rangle_A\rightarrow
 -{\sum e_q^2\over 16\pi^2}
 \Big({G^2(0)\over q^2}+c\alpha_s{G^3(0)\over q^4}+...\Big) 
\label{fla21}
\end{eqnarray}
where 
$G^2\equiv 2{\rm Tr}G_{\xi\eta}G_{\xi\eta}$,
$G^3\equiv 2{\rm Tr}G_{\xi\eta}G_{\eta\sigma}G_{\sigma\xi}$, and $c$ is
an (unknown) constant. (The 
 coefficient in front of $G^3$ vanishes at the tree level\cite{gkub}). 
\begin{figure}[htb]
\mbox{
\epsfxsize=16cm
\epsfysize=3.5cm
\hspace{0cm}
\epsffile{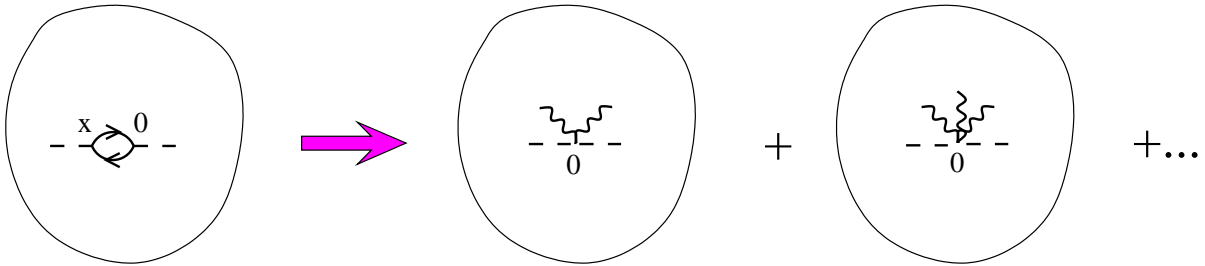}}
\vspace{0.5cm}
{\caption{ Expansion of the polarization operator 
in large-scale external fields. 
\label{fig3}}}
\end{figure}
Consider the leading term in this expansion;
since the field strength for the valley configuration (\ref{fla14}) depends
only on $x-x_0$
\begin{equation}
\int d^4x_0 {\rm Tr}G^v_{\xi\eta}(0)G^v_{\xi\eta}(0)=4S^v(z), 
\label{fla22}
\end{equation}
the intergal (\ref{fla20}) takes the form
\begin{eqnarray}
{1\over q^2}
\int_1^{\infty}dz\int_0^{\infty}{d\rho\over\rho^9} d^4R  
{1\over g^{17}(\rho)}F(z,{R^2/\rho^2})e^{-{S^v(z)\over g^2(\rho)}}
\label{fla23}
\end{eqnarray}
where we have included the factor  ${1\over 4\pi^2}\sum e_q^2S^v(z)$ in $F$. 
The (finite) integration over $R$ can be performed resulting in
an additional dimensional factor $\rho^4$: 
$$
\int d^4R F(\rho,z,R)=\rho^4\Phi(z)
$$ 
where the function
$\Phi$ is dimensionless so it can depend only on $z$. Finally, we get
\begin{equation}
{1\over q^2}\int_1^{\infty}dz\int_0^{\infty}{d\rho\over\rho^5}
{1\over g^{17}(\rho)}\Phi(z)  
e^{-{1\over g^2(\rho)}S^v(z)}.
\label{fla24}
\end{equation} 
Inverting 
Eq. (\ref{fla6}), we can write the corresponding contribution to Adler's
function as an
integral over the valley action ($t\equiv{S\over 16\pi^2}$):
\begin{eqnarray}
D(q^2)&\simeq&{1\over 3q^4}\int_0^{1}\!dt\!\int_0^{\infty}\!
{d\rho\over\rho^5} {1\over g^{17}(\rho)}\Phi(t) 
e^{-{4\pi\over \alpha_s(\rho)}t}.
\end{eqnarray}
Note that extra ${1\over g^2(\rho)}$ in the numerator can be eliminated using
integration by parts:
\begin{eqnarray}
&&{16\pi^2\over g^2(\rho)}\int_0^{1}\!dt e^{-{4\pi\over
\alpha_s(\rho)}t}\Phi(t)
=\int_0^{1}\!dt\Phi'(t)e^{-{4\pi\over\alpha_s(\rho)}t}
-\Phi(1)e^{-{4\pi\over\alpha_s(\rho)}}+\Phi(0).  
\end{eqnarray}
The last two terms are irrelevant for the would-be renormalon singularity 
at $t={2\over b}$ (the term $\sim
e^{-{4\pi\over\alpha_s(\rho)}}$ corresponds to the $I\bar{I}$ singularity 
and the term $\sim \Phi(0)$ does not depend on coupling constant).
Likewise, extra $g(\rho)$ can be absorbed by Laplace transformation
\begin{eqnarray}
&&
g(\rho)\int_0^{1}\!dt\Phi(t)e^{-{4\pi\over \alpha_s(\rho)}t}
=4\sqrt{\pi}\int_0^{1}\!dte^{-{4\pi\over \alpha_s(\rho)}t}
\int_0^t dt'(t-t')^{-1/2}\Phi(t')
\end{eqnarray}
leading to
\begin{eqnarray}
D(q^2)={1\over q^4}\int_0^{1}\!dt\!\int_0^{\infty}\!
{d\rho\over\rho^5}
e^{-{4\pi\over \alpha_s(\rho)}t}\Psi(t)
\label{fla25}
\end{eqnarray} 
where 
$\Psi(t)={1\over 3\sqrt{\pi}(4\pi)^{17}}\int_0^t dt'(t-t')^{-1/2}\Phi^{(9)}(t')$ 
after nine integrations by parts and a Laplace transformation.
Using the two-loop formula for $\alpha_s$
we get 
\begin{eqnarray}
D(t)\simeq\Psi(t)\int_0^{\infty}\!{d\rho\over q^4\rho^5}  
(q^2\rho^2)^{bt}\left({\alpha_s(\rho)/\alpha_s(q)}\right)^{{2b't\over b}}
\label{fla26}
\end{eqnarray} 
where $b'=51-{19\over 3}n_f$. It is clear now that the 
integral over $\rho$  
diverges in the IR region starting from $t={2\over b}$
\footnote{
Strictly speaking, at finite $q$ we cannot get to the singularity since
our semiclassical approach is valid up to $\rho<\Lambda_{QCD}$ which
translates to ${2\over b}-t>{1\over\ln q\Lambda_{QCD}}$; therefore we must 
take the limit $q^2\rightarrow\infty$ as well.
} leading to in a singularity in $D(t)$ at $t={2\over b}$. This singularity is
the first IR renormalon. At the one-loop level we can drop the ratio 
$\left({\alpha_s(\rho)/\alpha_s(q)}\right)^{{2b't\over b}}$ so the renormalon
is a simple pole ${1\over t-2/b}$. To get he character of this
singularity at the two-loop level, we recall that an extra
$\alpha(\rho)$ factor does not affect the singularity while an extra ${1\over
\alpha(q)}$ factor changes it from $(t-{2\over b})^\lambda$ to
$(t-{2\over b})^{\lambda-1}$, as one can easily  see from the integration by
parts \footnote{
To avoid confusion, note that in proving that extra 
${1\over\alpha_s(\rho)}$ does not change the singularity, we used the fact that
before the $\rho$ integration the function $\Phi(t)$, which is constructed from 
valley determinants, can be singular only at $t\rightarrow 1$ where these
determinants acquire zero modes.}:
\begin{eqnarray}
&&{16\pi^2\over g^2(q)}\int_0^{1}\!dt e^{-{4\pi\over
\alpha_s(q)}t}\left(t-{2\over b}\right)^\lambda
=\int_0^{1}\!dt\lambda\left(t-{2\over b}\right)^{\lambda-1}
e^{-{4\pi\over\alpha_s(q)}t}
+... 
\end{eqnarray}
Therefore, the extra factor
$\left({\alpha_s(\rho)/\alpha_s(q)}\right)^{{2b't\over b}}$ will convert
the simple pole ${1\over t-2/b}$ into a 
branching point singularity 
$(t-{2\over b})^{-1-4{b'\over b^2}}$\cite{mueller}. 

It is easy to see that the second term in the expansion (\ref{fla21}) gives the 
second renormalon 
singularity located at $t={3\over b}$. Higher terms of the expansion of the 
polarization operator (\ref{fla21})  will give the subsequent IR renormalons 
located at $t={4\over b},{5\over b},{6\over b}$ etc.
 
\section{UV renormalon as a dilatation mode}
 Next we demonstrate that the divergence of the integral (\ref{fla20}) 
 at small $\rho$ leads to the UV renormalon. In order to find the 
 polarization operator in the background of a very small vacuum fluctuation
(\ref{fla14})  we recall that this fluctuation is an inversion
\footnote{To get the Eq. (\ref{fla14}) literally, this inversion must be
accompanied by the gauge rotation with the matrix
$x(x-R)R/\sqrt{x^2(x-R)^2R^2}$, see ref. \cite{bshafer}.
}
\begin{eqnarray}
&&A^v_{\mu}(x-x_0)={\rho^2\over (x-x_0)^2}
\left(\delta_{\mu\alpha}-2{(x-x_0)_{\mu}(x-x_0)_{\alpha}\over y^2}\right) 
A^s_{\alpha}\Big({\rho^2\over (x-x_0)^2}(x-x_0)\Big)
\label{dop1}
\end{eqnarray}
of the
spherical configuration (\ref{fla13}) (gauge rotated by $x/\sqrt{x^2}$)
\begin{eqnarray}
&&A^s_{\alpha}(x)=\nonumber\\
&&~-i[\sigma_\alpha(\bar{x}-\bar{a})-
(x-a)_\alpha]\left({1\over (x-a)^2+\rho^2\xi}+
{\rho^2/\xi\over(x-a)^2((x-a)^2+\rho^2/\xi}\right),
\label{dop2}
\end{eqnarray}
where $a=R(\xi-1/\xi)$ (we chose $d=\rho$ for the inversion).
The corresponding transformation of the polarization operator has the form
\begin{eqnarray}
&&\Pi^v(x)=\left(\delta_{\mu\alpha}-2{y_{\mu}y_{\alpha}\over
y^2}\right)\langle j_{\alpha}\big({\rho^2\over y^2}y\big)
j_{\beta}\big(-{\rho^2\over x_0^2}x_0\big)\rangle_{A_s}
\left(\delta_{\beta\mu}-2{x_{0\beta}x_{0\mu}\over x_0^2}\right)
\label{dop3}
\end{eqnarray}
where we use the notation $y\equiv x-x_0$. In the limit
$\rho\rightarrow 0$ we need the polarization operator in the background of
$A^s$ at small distances (see Fig. 4),
\begin{figure}[htb]
\mbox{
\epsfxsize=14cm
\epsfysize=4.5cm
\hspace{1cm}
\epsffile{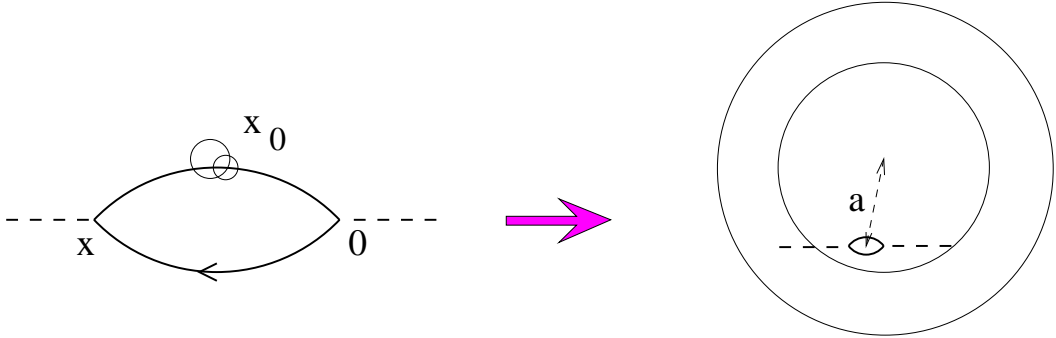}}
\vspace{0cm}
{\caption{ Inversion of the polarization operator in the valley
background. 
\label{fig4}}}
\end{figure}
so we can use the expansion (\ref{fla21}) (in the
coordinate space)
\begin{eqnarray}
&&\langle j_{\mu}(x)j_{\nu}(0)\rangle_A~\rightarrow~-{\sum e_q^2\over 384\pi^4}
 \Big[{2x_{\mu}x_{\nu}+x^2\delta_{\mu\nu}\over x^4}G^2(0)+
 c\alpha_sG^3(0)
 \big(2{x_{\mu}x_{\nu}\over x^2}-3\delta_{\mu\nu}\ln x^2\big)+...\Big]
\label{fla21a}
\end{eqnarray}
and obtain
\begin{eqnarray}
&&\Pi^v(x)\rightarrow\nonumber\\
&&
{\sum e_q^2\over 192\pi^4}
 \Bigg[{\rho^8G_s^2(0)\over y^4x_0^4x^2}
\Big\{1-{4(x_0y)^2\over x_0^2y^2}\Big\}+
c\alpha_s{\rho^{12}G_s^3(0)\over y^6x_0^6}
\Big\{
\left(3\ln{\rho^2x^2\over y^2x_0^2}-1\right){2(x_0y)^2\over x_0^2y^2}
+1\Big\}\Bigg]
\label{fla27}
\end{eqnarray}
where $G^s_{\mu\nu}(0)$ is the field strength of the spherical configuration 
(\ref{dop2}) calculated at
at the origin.  
After the integration over $x$ and $x_0$ the first term $\sim G_s^2$ vanishes 
and the second gives
\begin{eqnarray}
&&\int dx e^{iqx} \int dx_0{1\over y^6x_0^6}\Big[
\Big(3\ln{x^2\over y^2x_0^2}-1\Big){2(x_0y)^2\over x_0^2y^2}
+1\Big]={\pi^4\over 32}(6\ln\pi+3C -2)
q^4\ln^2 q^2 \,,
\label{fla27a}
\end{eqnarray} 
leading to
\begin{eqnarray}
&&\int dx dx_0 e^{iqx} \Pi^v(x)=
c'\alpha_s(q){\sum e_q^2\over 64}\rho^{6}{\cal G}(z,R^2/\rho^2)
q^4\ln^2 q^2
\label{fla28}
\end{eqnarray} 
where
\begin{equation}
{\cal G}(z,R^2/\rho^2)=\rho^{6}G_s^3(0)
\label{fla29}
\end{equation} 
is a dimensionless non-singular function of $z$ and $R^2/\rho^2$
(the explicit expression can be easily found from Eq. (\ref{dop2})). 
The argument of $\alpha_s$ in Eq. (\ref{fla27}) is determined by the
characteristic momenta  in the loop with one extra gluon in the valley
background. After inversion, the characteristic distances in the loop 
diagram determining
the coefficient in front of $G_s^3$ are 
$\sim \rho^2\sqrt{{x^2\over x_0^2y^2}}$ which
means that before inversion the characteristic momenta in the loop 
were $\sim q$ (in the integral (\ref{fla30})
 $x^2\sim x_0^2\sim y^2\sim q^{-2}$).

Performing  the integration over $R$
we obtain the analog of the Eq. (\ref{fla25}) for the UV renormalon
\begin{equation}
D(q^2)\simeq q^2\alpha_s(q)\int_0^{1}\!dt\!\int_0^{\infty}\!
d\rho \rho (\ln q^2\rho^2)^2 
e^{-{4\pi\over \alpha_s(\rho)}t}\tilde{\Psi}(t)
\label{fla30}
\end{equation} 
leading to 
\begin{eqnarray}
D(t)\simeq
\tilde{\Psi}(t)q^2\!\int_0^{\infty}\!\!d\rho^2 
(q^2\rho^2)^{bt}(\ln q^2\rho^2)
\left(\alpha_s(\rho)/\alpha_s(q)\right)^{{2b't\over
b}}.
\label{fla31} \end{eqnarray} 
Note that extra $\alpha_s(q)$ in  Eq. (\ref{fla30}) is compensated by one 
power of
$\ln q^2\rho^2$. The integral over $\rho$ diverges at $t=-{1\over b}$ which 
is the position 
of the first UV renormalon. At the one-loop level this renormalon is a 
double pole $(1+t/b)^{-2}$, same as in the perturbative analysis
\cite{beneke}). Subsequent terms in the
expansion (\ref{fla21})  correspond to the UV renormalons located at
$t=-{2\over b}, -{3\over b},-{4\over b}...$ 
\footnote{
We do not
see the singularity at 
$t={1\over b}$ which is proposed in \cite{zakharov}.
However, this singularity is due to the
small-size monopoles which are beyond the scope of this
paper (our
results are based on the gaussian integration near the finite-action 
vacuum fluctuations while monopoles have infinite action).}. 
It should be mentioned that the Eq. (\ref{fla31}) cannot reproduce
the strength of the first UV renormalon at the two-loop level\cite{kivel}. The
reason is that in  Eqs. (\ref{fla21}), (\ref{fla27}) we have 
neglected the anomalous dimensions of the operators 
$\sim(\alpha_s(q)/\alpha_s(\rho))^{\gamma\over b}$. Such factors can change the
strength of the singularity. For the IR renormalon this does not matter since
the operator  $G^2$ is renorm-invariant ($\gamma=0$) and for the subsequent
renormalons  we can easily correct our results by corresponding $\gamma$'s.
Unfortunately, for the UV renormalons we do not know how to use the conformal
invariance with the anomalous dimensions included.

\section{Conclusions}
We have demonstrated that the integration along the dilatation modes 
in the functional space near the $I\bar{I}$ configuration leads to 
the renormalon singularities in the Borel plane.  
It is very important to note that we actually never use the explicit form of
the valley configuration. What we have really used are the three facts:
(I) conformal anomaly $\Rightarrow$ the 
fact that the rescaling of the vacuum fluctuation with an action S by a 
factor $\lambda$ multiplies the determinant by $\lambda^{bS/8\pi^2}$ 
leading to the formula (\ref{fla11}),
(II) the expansion of the polarization operator (\ref{fla21}) 
in slow varying fields, and   
(III) the conformal invariance of QCD at the 
tree level (for the UV renormalon we wrote down the small-size 
configuration as an inversion of a large-scale vacuum fluctuation). 
All of these properties hold true for an arbitrary vacuum fluctuation so 
we could take an arbitrary valley 
and arrive at the same results Eq. (\ref{fla26}) and 
Eq. (\ref{fla31})
\footnote{ 
The only difference would be that the upper limit 
for the integration over $t$ will be $\infty$ rather than 1 
because the arbitrary valley starts at the perturbative vacuum and 
goes to the infinity with constantly increasing action
.}.
It means that 
our result about the renormalon singularity
 coming from the  dilatation mode in the functional space is general.
 
The form of the polarization operator in a valley background  
suggests the following parametrization of Adler' function as a
double integral in $S$ and $\rho$ 
\begin{equation}
D(\alpha_s(q))=\int_0^{\infty}dt\int
{d\rho^2\over\rho^2}d(q^2\rho^2,t)e^{-{4\pi\over\alpha_s(\rho)}t}
\label{br1}
\end{equation}
where 
\begin{equation}
d(q^2\rho^2,t)\sim (q^2\rho^2)^{-2} ~~{\rm at}~~\rho\rightarrow\infty,
~~~~~~~~~~~~d(q^2\rho^2,t)\sim (q^2\rho^2) ~~{\rm at}~~\rho\rightarrow 0
\label{br2}
\end{equation}
The function $d(q^2\rho^2,t)$ contains only instanton-induced
singularities in $t$ whereas the IR and UV renormalons come from the divergence
of the integral (\ref{br1}) at large or small $\rho$, respectively. 
If we
``switch off'' the running coupling constant (e.g consider the conformal
theory without $\beta$-function) the  representation of Adler's function takes
the form 
\begin{equation}
D^{\rm conf}(\alpha_s)=\int_0^{\infty}dte^{-{4\pi\over\alpha_s}t}D^{\rm
conf}(t) \label{br3}
\end{equation}
where $D^{\rm conf}(t)=\int {d\rho^2\over\rho^2}d(q^2\rho^2,t)$ 
(the integral over $\rho$ converges due to Eq. (\ref{br2})). 
In real QCD the function $D^{\rm conf}(\alpha_s)$ defines the ``conformal
expansion''  of the Adler's function with
the coefficients coming from the ``skeleton diagrams'' in terms
of usual perturbation theory\cite{brodsky}. 

As we mentioned above, our analysis of renormalons is applicable 
to the ``off-shell'' processes which can be related to the Euclidean 
correlation functions of two (or more) currents. One of possible 
directions of the future work would be to generalize these ideas to
the renormalons in the ''on-shell'' (Minkowskian) processes intensively 
discussed in the current literature (see discussion in ref. \cite{beneke}). 

Finally, we note that our method gives same results as the conventional
approach - namely position and strength  of the renormalon singularity, 
but not the coefficient in front of it. The coefficient in front of the first 
IR renormalon is
expremely important since it determines the  numerical value of asymptotics of
perturbative series for $R_{e^+e^-\rightarrow{\rm hadrons}}$.  In principle,
for a given valley it is possible to calculate this coefficient in the first
order in coupling constant $g^2(\rho)$ since it is given by a  product of the 
determinants in the valley background, which can at least be computed
numerically. 
However, the fact that an extra $g^2(\rho)$ does not change the
position and/or character of the singularity means that all terms in the 
perturbative series
in $g^2(\rho)$ contribute on equal footing. This is closely related to the 
fact that the choice of the valley is not unique: after the change of the
valley, the new leading-order coefficient, which comes from the determinants in
the  background of a new configuration,
is given by an infinite perturbative 
series in $g^2(\rho)$ (coming from quantum corrections) 
in terms of the original valley.
As a result, finding the coefficient in front of the leading IR singularity 
would require the integration over all possible valleys.  
This study is in progress.

\vskip0.5cm
\paragraph*{Acknowledgements.}
One of us (I.B.) is grateful to V.M. Braun, S.J. Brodsky, R.L. Jaffe, 
A.H. Mueller, and E.V. Shuryak for valuable 
discussions. 
This work was supported by the US Department of Energy under contracts 
DE-AC05-84ER40150 and DE-FG02-97ER41028.
\section{Appendix}
We will prove  
that rescaling of the arbitrary configuration by factor
$\lambda$ (so that $\rho\rightarrow \lambda\rho$)
leads at the one-loop level to multiplication of the determinant 
by a factor of
\begin{equation}
\lambda^{bS/8\pi^2}
\label{xz0}
\end{equation}
where $S$ is the action of this configuration. 
Let us demonstrate it for the 
determinant of
the Dirac operator in the background of a field 
$A_{\mu}(x;\rho)$ characterized by the size $\rho$.
We will prove that
\footnote{
 For simplicity, we
assume that the Dirac operator has no zero modes, as in the case of 
the $I\bar{I}$ valley (for the treatment of Dirac operator with zero
modes see e.g. \cite{shvarz}).} 
\begin{equation}
\sqrt{\det \not\! \tilde{P}^2}=\sqrt{\det \not\! P^2}~
\lambda^{{S\over 12\pi^2}}
\label{xz1}
\end{equation}
where $P_{\mu}=i\partial_{\mu}+A_{\mu}(x)$ is the operator of the 
covariant momentum for our configuration and
$\tilde{P}_{\mu}=i\partial_{\mu}+\tilde{A}_{\mu}(x)$ is the 
covariant momentum in the background of the stretched configuration
$\tilde{A}_{\mu}(x)=A_{\mu}(x;\lambda\rho)=
\lambda^{-1}A_{\mu}(\lambda^{-1}x;\rho)$. Using Schwinger's
notations we can write down
\begin{equation}
\ln\det \not\! \tilde{P}^2=-\int_0^{\infty}{ds\over s}
\int d^4x {\rm Tr}(x|e^{-s\not \tilde{P}^2}|x)
\label{xz2}
\end{equation}
where $|x)$ are the eigenstates of the coordinate operator normalized according
to $(x|y)=\delta^{(4)}(x-y)$. For the Dirac operator,
the integral in Eq. (\ref{xz2}) diverges so we will assume a cutoff 
$s>\epsilon$ and take $\epsilon\rightarrow 0$ in the final results. 
After change of variables
$x\rightarrow\lambda^{-1} x$, the integral in the r.h.s. of Eq. (\ref{xz2}) 
reduces to 
\begin{equation}
\ln\det \not\! \tilde{P}^2=-\int_0^{\infty}{ds\over s}
\int d^4x {\rm Tr}(x|e^{-s\lambda^{-2}\not{P}^2}|x)
\label{xz3}
\end{equation}
so
\begin{eqnarray}
&&\ln\det \not\! \tilde{P}^2-\ln\det \not\! P^2=\nonumber\\
&&
\int_0^{\infty}{ds\over s}
\int d^4x {\rm Tr}(x|e^{-s\lambda^{-2}\not
{P}^2}-e^{-s\not{P}^2}|x)=-\lim_{\epsilon\rightarrow
0}  \int^{\epsilon}_{\lambda^{-2}\epsilon}{ds\over s}\int d^4x {\rm
Tr}(x|e^{-s\not{P}^2}|x) 
\label{xz4}
\end{eqnarray}
Using the well-known result for the DeWitt-Seeley coefficients for the 
Dirac operator (see e.g. \cite{dewitt})
\begin{equation}
{\rm Tr}(x|e^{-s\not{P}^2}|x)=
{3\over 4\pi^2s^2}+{1\over 48\pi^2}G^a_{\mu\nu}(x)G^a_{\mu\nu}(x),
\label{xz5}
\end{equation}
we obtain 
\begin{equation}
\ln\det \not\! \tilde{P}^2=\ln\det \not\! P^2-{\ln\lambda\over 24\pi^2}\int
d^4x  G^{va}_{\mu\nu} G^{va}_{\mu\nu} 
\label{xz6}
\end{equation}
which corresponds to the Eq. (\ref{xz1}). In the case of $n_f$ quark flavors,
the coefficient ${1\over 12\pi^2}$ in r.h.s. of Eq. (\ref{xz1}) will 
be multiplied by $n_f$ leading to the factor 
$\lambda^{-{2\over3}n_fS/8\pi^2}$. This factor is easily
recognized as a quark part of the one-loop coefficient $b$ for Gell-Mann-Low
$\beta$-function in Eq. (\ref{xz0}). Similarly, the rescaling of the gluon (and
ghost) determinants reproduces $11$ - the gluon part of the one-loop
coefficient $b$ in Eq. (\ref{xz0}). Thus we have shown that $e^{-{S\over
g^2}}\rightarrow e^{-{S\over g^2(\rho)}}$ with one-loop accuracy. Using the
two-loop formulas for the Seeley coefficients one could prove that $e^{-{S\over
g^2(\rho)}}$ reproduces at the two-loop level and  demonstrate that
$g^2\rightarrow g^2(\rho)$ in the pre-exponential factor $g^{-17}$ as well 
(as it follows from the renormalizability of the theory).

\end{document}